\documentclass{article}
\usepackage{spconf,amsmath,graphicx}
\usepackage{amsfonts}
\usepackage{epsfig,amssymb,latexsym,graphics,float}
\usepackage{setspace,subfig}
\usepackage{citesort}
\usepackage{amsopn}
\usepackage{amssymb,times}
\usepackage[ruled,vlined,linesnumbered]{algorithm2e}
\usepackage{ifpdf}

\title{ILAPF: Incremental Learning Assisted Particle Filtering}
\name{Bin~Liu$^{\star}$
\thanks{$^\star$Address correspondence to bins@ieee.org. This work was partly supported by the National Natural Science Foundation (NSF) of
China (Nos. 61571238 and 61572263), the China Postdoctoral Science Foundation (Nos. 2015M580455 and 2016T90483), the Six Talents Peak
Foundation of Jiangsu Province under grant No. XYDXXJS-CXTD-006 and Nanjing University of Posts and Telecommunications Yancheng Big Data Research Institute.}}
\address{School of Computer Science, \\
Jiangsu Key Lab of Big Data Security $\&$ Intelligent Processing, \\
Nanjing University of Posts and Telecommunications, Nanjing, 210023 China}
\begin{document}
\maketitle
\begin{abstract}
This paper is concerned with dynamic system state estimation based on a series of noisy measurement with the presence of outliers.
An incremental learning assisted particle filtering (ILAPF) method is presented, which can learn the value range of outliers incrementally
during the process of particle filtering. The learned range of outliers is then used to improve subsequent filtering of the future state.
Convergence of the outlier range estimation procedure is indicated by extensive empirical simulations using a set of differing
outlier distribution models. The validity of the ILAPF algorithm is evaluated by illustrative simulations,
and the result shows that ILAPF is more accurate and faster than a recently published state-of-the-art robust particle filter.
It also shows that the incremental learning property of the ILAPF algorithm provides an efficient way
to implement transfer learning among related state filtering tasks.
\end{abstract}
\section{Introduction}\label{sec:intro}
State filtering plays a key role in the field of signal processing. This paper is concerned with nonlinear state filtering,
for which the particle filter (PF) has been widely accepted as a well-established methodology for use \cite{arulampalam2002tutorial,carpenter1999improved,gordon1993novel,liu2010multi,liu2013sequential,liu2008particle}. For conventional PF methods,
a major degradation in performance will happen when a significant mismatch between the leveraged model and the real mechanism
that governs the system's evolution exists. A popular strategy to handle such issue of model mismatch is to
employ a set of candidate models, instead of a single model, to take account of model uncertainty.
To this end, a number of multiple-model strategies (MMS) based PF algorithms have been proposed in the
literature \cite{liu2017robust,yi2016robust,liu2011instantaneous,Drovandi2014,Inigo2016sequential}.

In this paper, we consider the problem of nonlinear state filtering in presence of outliers.
Instead of resorting to commonly used MMS that usually specify a set of candidate models beforehand to
take account of model uncertainty, we select to learn an approximation of
the outlier distribution in a sequential way. Specifically, we approximate the outlier distribution by a simple uniform
distribution and then use an outlier range estimation (ORE) procedure to estimate the lower and upper bounds of that distribution.
Convergence of the ORE procedure is indicated by extensive simulations as shown in Subsection \ref{sec:ore}.
We incorporate the developed ORE operations into each iteration of PF, referring the resulting algorithm as Incremental Learning Assisted PF (ILAPF).
This algorithm is shown to be robust against outliers, more accurate and faster than a robust PF method
published in ICASSP 2017 \cite{liu2017robust}, and can provide an efficient way to implement transfer learning among related state filtering tasks.
\section{Model}\label{sec:model}
In this Section, we present a succinct description  of the model we use, based on which the ILAPF algorithm is developed.
Following \cite{liu2017robust}, we consider a state space model as follows
\begin{eqnarray}
x_k&=&f(x_{k-1})+u_k\\
y_k&=&h(x_k)+n_k,
\end{eqnarray}
where $k$ denotes the time index, $x\in\mathbb{R}^{d_x}$ denotes the state of the dynamic system to be estimated,
$y\in\mathbb{R}^{d_y}$ the measurement of $x_k$, $f$ the state transition function, $h$ the measurement function,
$u$ the independent identically distributed (i.i.d.) process noise and $n$ the i.i.d. measurement noise.
In a classical problem setting, the probability density functions (pdfs) of $u_k$ and $n_k$ are assumed \emph{a priori} known,
which determine the state transition prior density $p(x_k|x_{k-1})$ and the likelihood function $p(y_k|x_k)$, respectively.
Then using Bayesian theorem, we can formulate the state filtering problem as the computation of the \emph{a posteriori}
pdf of $x_k$ given $y_{1:k}\triangleq\{y_1,\ldots,y_k\}$, denoted as $p(x_k|y_{1:k})$ (or in short $p_{k|k}$). Recursive solutions exist
since $p_{k|k}$ can be computed from  $p_{k-1|k-1}$ recursively as follows
\begin{equation}\label{eqn:filter}
p_{k|k}=\frac{p(y_k|x_k)\int p(x_k|x_{k-1})p_{k-1|k-1}dx_{k-1}}{p(y_k|y_{1:k-1})}.
\end{equation}

Here we bring a variable $o\in\{0,1\}$ to take account of the uncertainty in the measurement model.
Specifically, let $o_k=1 (0)$ denote the event that $y_k$ is (is not) an outlier.
If $o_k=0$ we assume that the measurement noise $n_k$ is Gaussian distributed by default, namely $n_k\sim\mathcal{N}(0,R)$,
where $R$ is \emph{a priori} known.
If $o_k=1$ we assume that $n_k$ is generated from an unknown uniform
distribution $\mathcal{U}(lb,ub)$, where $lb$ and $ub$ denote the lower and upper bounds of $\mathcal{U}$, respectively.
The likelihood function can now be represented as follows
\begin{equation}
p(y_k|x_k)=\left\{\begin{array}{ll}
p(y_k|x_k,o_k=1),\quad\mbox{if}\quad o_k=1 \\
p(y_k|x_k,o_k=0),\quad\mbox{if}\quad o_k=0 \end{array} \right.
\end{equation}
where
\begin{eqnarray}\label{eqn:likelihood_outlier}
p(y_k|x_k,o_k=1)&=&\left\{\begin{array}{ll}
1/V_{lb,ub},\quad\mbox{if}\quad e_k\in[lb,ub] \\
0,\quad\mbox{otherwise} \end{array} \right.\\
p(y_k|x_k,o_k=0)&=&\mathcal{N}\left(e_k|0,R\right),
\end{eqnarray}
where $e_k=y_k-h(x_k)$ and $V_{lb,ub}$ denotes the volume of the space bounded by $lb$ and $ub$.
An ORE procedure is developed to estimate $lb$ and $ub$ incrementally, see details in Subsection \ref{sec:ore}.

\section{The proposed ILAPF Algorithm}
Here we present the ILAPF algorithm used to address the Bayesian state filtering problem defined by Eqns.(1)-(6).
Suppose that, at time step $k-1$ ($k>1$), we have a discrete approximation of $p(x_{1:k-1}|y_{1:k-1})$ given by a set of weighted
samples $\{x_{1:k-1}^i,\omega_{k-1}^i\}_{i=1}^N$, $\sum_{i=1}^N\omega_{k-1}^i=1$.
At time $k$, the $i$th sample is first extended by a particle $\hat{x}_k^i=f(x_{k-1}^i)$.
Then, according to importance sampling theory \cite{doucet2000sequential,arulampalam2002tutorial,smith2013sequential}, it
is weighted by
\begin{equation}\label{eqn:weight}
\omega_{m,k}^i=\omega_{k-1}^ip(y_k|\hat{x}_k^i,o_k=m),
\end{equation}
under the hypothesis $o_k=m$, $m=0,1$. Then the likelihood of the event $o_k=m$ is given by
\begin{equation}
L(o_k=m)=\sum_{i=1}^N\omega_{m,k}^i.
\end{equation}
Assume that the prior probability $p(o_k=0)=p(o_k=1)=0.5$. From Bayesian theorem the posterior probability of $o_k=m$ is given by
\begin{equation}\label{eqn:post_prob_outlier}
\pi(o_k=m)=\frac{L(o_k=m)}{L(o_k=0)+L(o_k=1)}, m=0,1.
\end{equation}
Then the importance weights of the particles can be calculated as follows
\begin{equation}\label{eqn:weight_update}
\omega_{k}^i\varpropto\sum_{m=\{0,1\}}\pi(o_k=m)\tilde{\omega}_{m,k}^i, i=1,\ldots,N,
\end{equation}
where $\tilde{\omega}_{m,k}^i=\omega_{m,k}^i/\sum_{j=1}^N\omega_{m,k}^j$.
To get around of particle degeneracy, we use a resampling step to discard
the particles with low weights and duplicate those with high
weights. For details on resampling techniques used by PF methods, readers are referred to \cite{douc2005comparison,Li2015Resampling,Hol2006on}.

The above operations constitute the major building block of the ILAPF algorithm, while a crucial issue is neglected, that is how to compute
$p(y_k|\hat{x}_k^i,o_k=1)$ in Eqn. (\ref{eqn:weight}) with an unknown outlier distribution.
We present the ORE procedure in Subsection \ref{sec:ore}
to address the above issue, and then summarize the operations of ILAPF in Subsection \ref{sec:operation_ilapf}.
\subsection{The outlier range estimation (ORE) procedure}\label{sec:ore}
Assume that the whole population of the outliers has a definite value range specified by a lower and upper bounds $lb$
and $ub$. We can estimate $lb$ and $ub$ accurately provided that we have enough outlier data points at hand.
But in practical tasks, usually, only a sparse set of outliers can be collected in a sequential way.
The question under consideration here is: how to estimate $lb$ and $ub$ accurately using a
limited number of outliers that have been found?
To fit the sequential structure of the state filtering problem,
we also expect that the estimation procedure can be performed in a sequential way.

\begin{figure}[t]
\centering
\includegraphics[width=1.75in,height=1.3in]{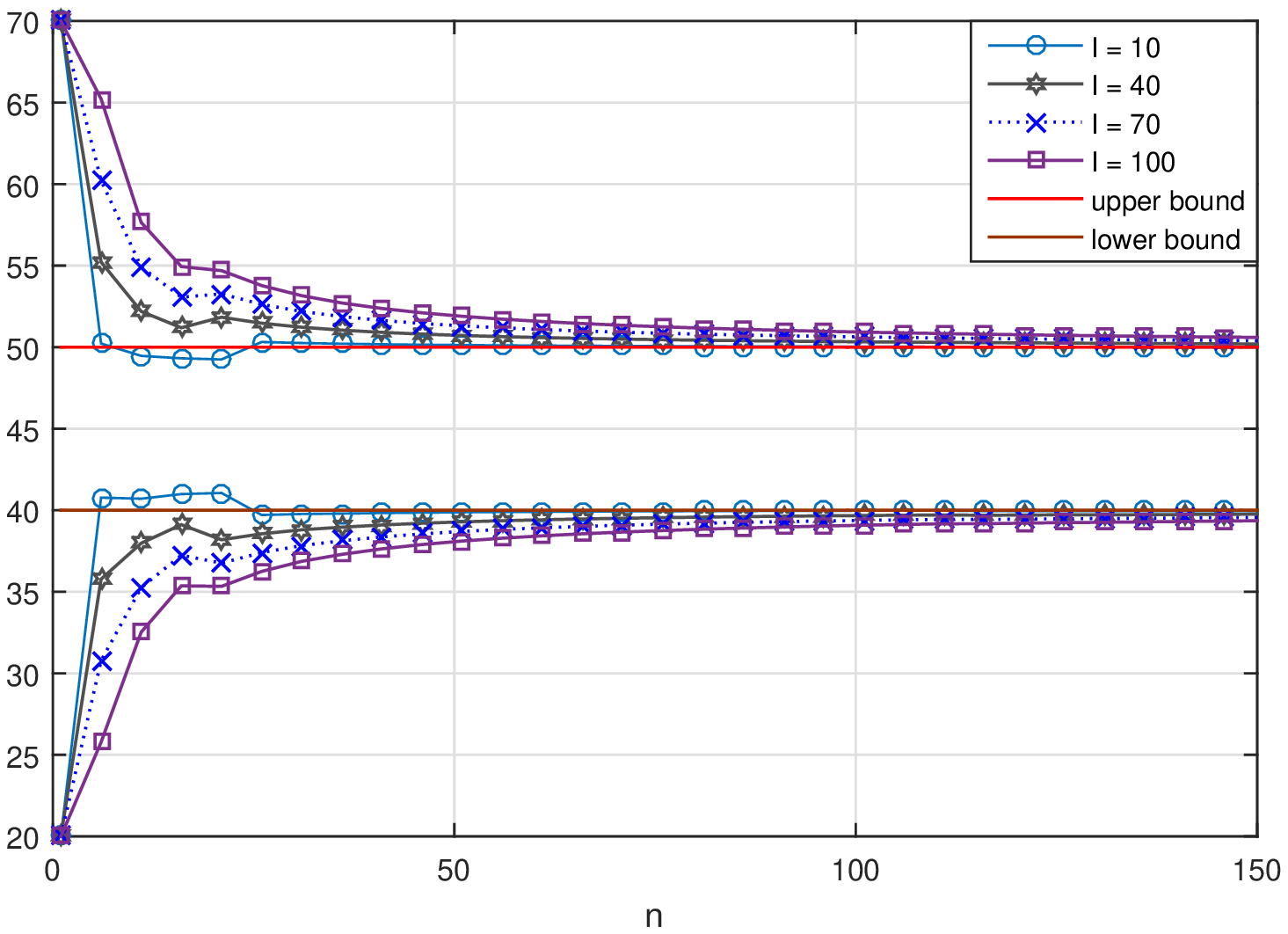}\includegraphics[width=1.75in,height=1.3in]{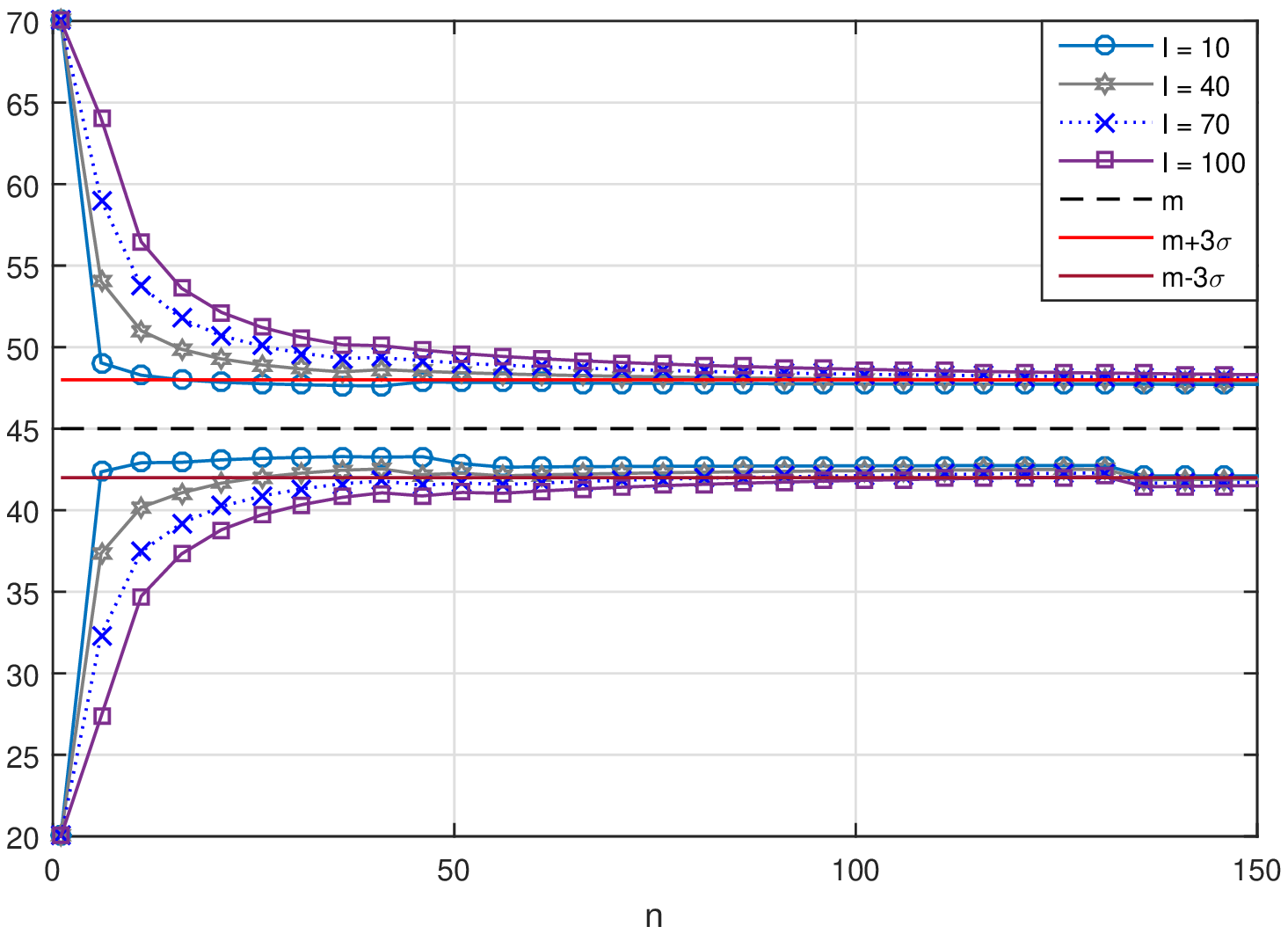}\\
\includegraphics[width=1.75in,height=1.3in]{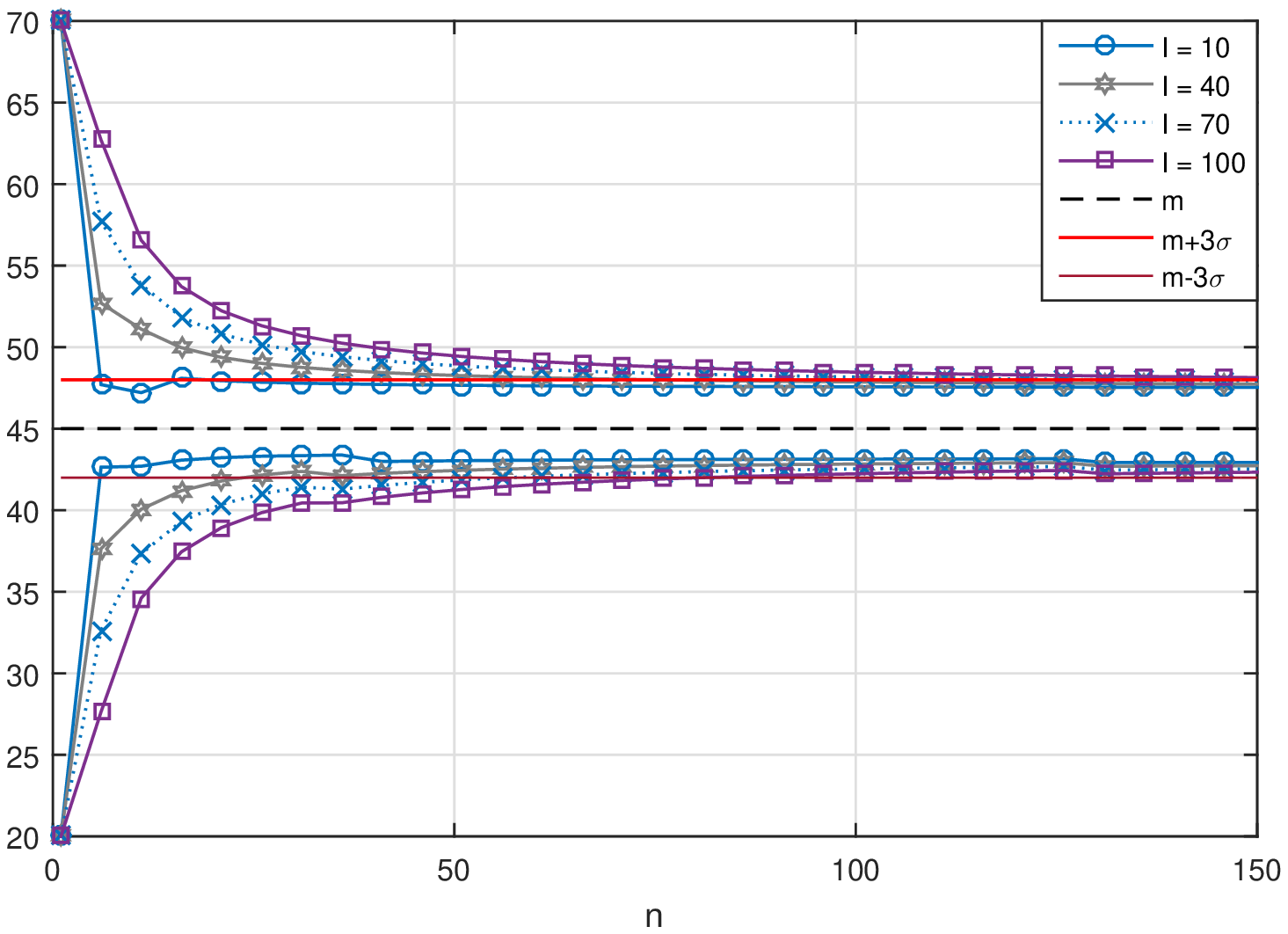}\includegraphics[width=1.75in,height=1.3in]{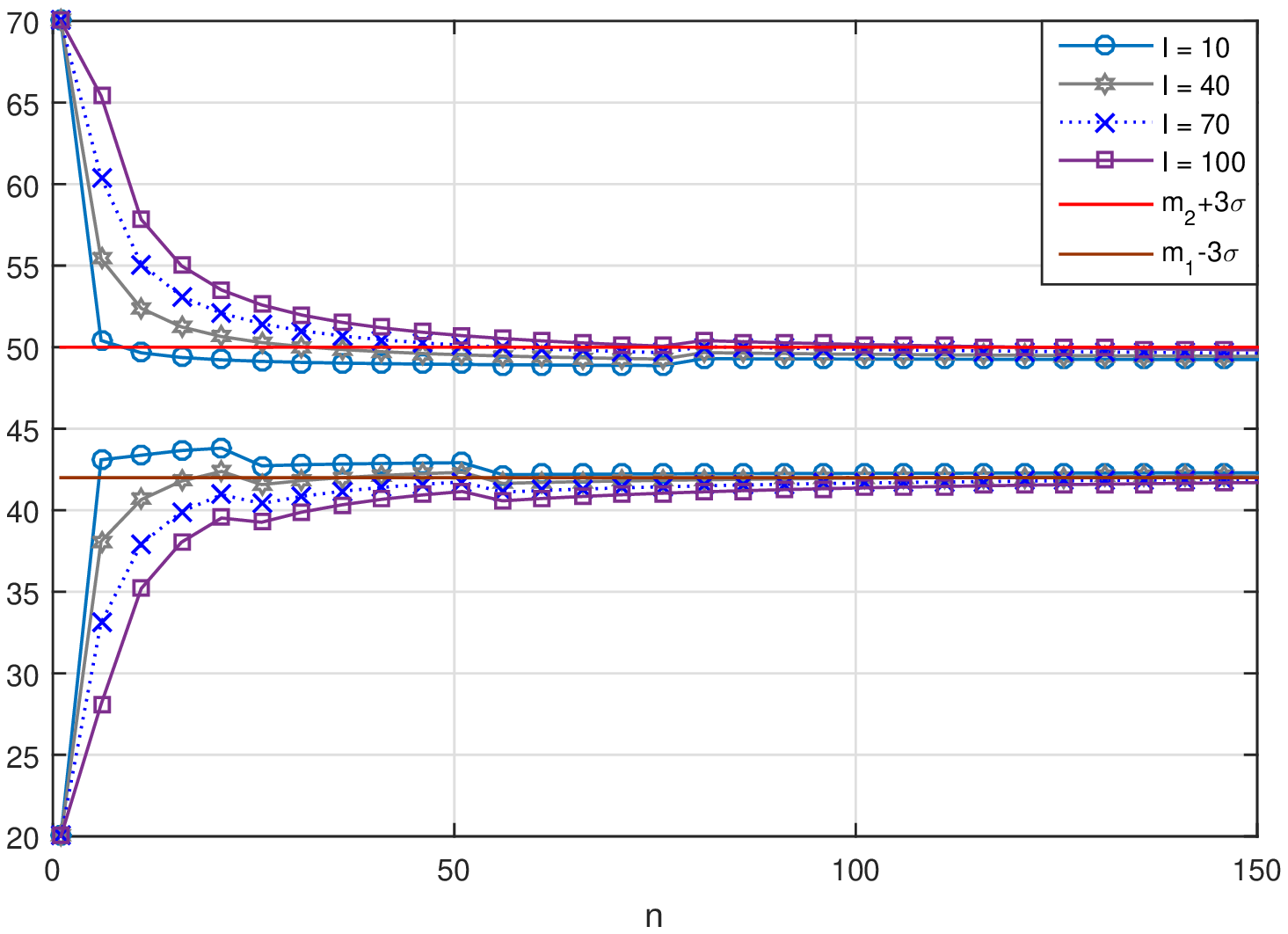}
\caption{The simulation result of using the ORE procedure to sequentially estimate the range of the outliers.
4 distributions to generate the outliers are considered, including the Uniform, Gaussian, Student's t and
a two-component Gaussian mixture distribution, corresponding to the top left, top right, bottom left and bottom right panel,
respectively. The X-coordinate represents the number of outliers that have been detected and the Y-coordinate the value of the outliers.
$m$ and $\sigma$ denote the mean and the standard error of a related distribution, respectively.
In the bottom right panel, $m_1$ and $m_2$ represent the smaller and bigger mean of the two Gaussian components that
have the same standard error $\sigma$.}\label{fig:ore}
\end{figure}

We develop an ORE method to address the above problem.
This method only has one free parameter $I$, which can be interpreted as a measure of uncertainty.
In ORE method, we consider outliers sequentially, making an incremental update to the estimation of $lb$ and $ub$,
once a new outlier arrives.
Assume that the current estimations of $lb$ and $ub$ are $\hat{lb}$ and $\hat{ub}$, respectively, and the number of
outliers that have been found is $n$. When the $(n+1)$th outlier, denoted as $z_{n+1}$, arrives, the ORE
procedure updates $\hat{lb}$ and $\hat{ub}$ as follows
\begin{eqnarray}\label{eqn:bound_update}
\hat{lb}&=&\min\{\hat{lb},z_{n+1}\}-I/(n+1),\\
\hat{ub}&=&\max\{\hat{ub},z_{n+1}\}+I/(n+1).
\end{eqnarray}
We tested the validity of the ORE method via simulations. We used 4 differing outlier distributions to simulate the outliers,
including a Uniform $\mathcal{U}(40,50)$, a Gaussian $\mathcal{N}(45,1)$, a Student's t and
a two-component Gaussian mixture distribution $0.5\mathcal{N}(45,1)+0.5\mathcal{N}(47,1)$.
The Student's t distribution has a degree of freedom 3, mean 45
and standard error 1. The ORE procedure is initialized with $\hat{lb}=20$ and $\hat{ub}=70$, which represent an initial guess of the bounds.
We considered 4 values of $I$, namely 10, 40, 70 and 100. For each $I$ value and each outlier distribution,
we ran the ORE method to process the data items that arrive one by one.
We recorded the estimated bounds at each time step when an outlier arrives, and the result is shown in Fig.\ref{fig:ore}.
We see that, for every distribution case, the estimated bounds converge to the desired ones and the convergence rate depends on the value of $I$.
Specifically, Fig.\ref{fig:ore} shows that an $I$ value between 10 and 40 will be best for choice.
So for the simulated experiments presented in Section \ref{sec:simu}, the value of $I$ is set at 20.
Note that we take mean$\pm3\times$standard error as the desired bounds for the Gaussian
and Student's t cases, and $m_1-3\sigma$ and $m_2+3\sigma$ for the Gaussian mixture case, where
$m_1$ and $m_2$ denote the smaller and bigger mean value, $\sigma$ standard error of those 2 mixing Gaussian components.

\subsection{A summarization of the operations in ILAPF}\label{sec:operation_ilapf}
Starting from $\{x_{k-1}^i,\omega_{k-1}^i\}_{i=1}^N$, $\hat{lb}$ and $\hat{ub}$ and the number of outliers
that have been found $n$, we present operations in one iteration of the ILAPF algorithm corresponding to time step $k$ as follows.
\begin{itemize}
\item Sampling step. Sample $\hat{x}_k^i$ from the state transition prior by setting $\hat{x}_k^i=f(x_{k-1}^i)$,
$i=1,\ldots,N$;
\item Weighting step. Set $\omega_{m,k}^i$ using Eqn.(\ref{eqn:weight}), $i=1,\ldots,N$.
Compute $\pi(o_k=1)$ and $\pi(o_k=0)$ using Eqn. (\ref{eqn:post_prob_outlier}).
Calculate importance weight $\omega_{k}^i$ using Eqn. (\ref{eqn:weight_update}), $i=1,\ldots,N$.
Then normalize these weights to guarantee that $\sum_{i=1}^N\omega_{k}^i=1$;
\item ORE step.
If $\pi(o_k=1)>0.5$, declare $y_k$ to be an outlier, let $n=n+1$ and update $\hat{lb}$ and $\hat{ub}$
using Eqns. (11) and (12), respectively. Note that $\pi(o_k=1)$ is an output of the above weighting step.
\item Resampling step. Sample $x_k^i\sim\sum_{j=1}^N\omega_k^j\delta_{\hat{x}_k^j}$,
set $\omega_k^i=1/N$, $i=1,\ldots,N$. $\delta_{x}$ denotes the Dirac-delta function located at $x$.
\end{itemize}
\section{Performance Evaluation}\label{sec:simu}
We evaluated the validity of ILAPF via illustrative simulations. A recently published
heterogeneous mixture model based robust PF (HMM-RPF) \cite{liu2017robust} was included for performance comparison.
\subsection{Simulation Setting}\label{sec:setting}
The simulation setting is similar as that in \cite{liu2017robust}.
We design a modified version of the time-series experiment as presented in \cite{van2000the} by replacing some normal measurements with outliers.
The time-series is generated as follows
\begin{equation}
x_{k+1}=1+\sin(0.04\pi\times(k+1))+0.5x_k+u_k, 1\leq k<60,
\end{equation}
with the value of $x_1$ set at 1, and $u_k$ represented as a Gamma(3,2) random variable modeling the process noise. The measurement model is
\begin{equation}\label{measure_func_simu}
y_k=\left\{\begin{array}{ll}
0.2x_k^2+n_k,\quad\quad\quad k\leq30 \\
0.2x_k-2+n_k,\quad\, k>30 \end{array} \right.
\end{equation}
The measurement noise $n_k$ is Gaussian distributed by default with mean 0 and variance 0.01.
The outliers arrive at time steps $k=7, 8, 9, 20, 37, 38, 39, 50$.
Each outlier is simulated by replacing the default Gaussian distribution with a uniform distribution $\mathcal{U}(20, 30)$ in generating
the value of $n_k$ in Eqn.(\ref{measure_func_simu}).
The state filtering algorithms to be tested are set to be blind to both the arrival time and the generative distribution of the outliers.
\subsection{Results}
\begin{figure}[t]
\centering
\includegraphics[width=3in,height=1.7in]{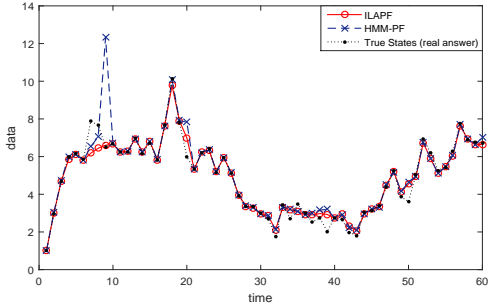}
\caption{Filtering result of ILAPF and HMM-RPF \cite{liu2017robust}.}\label{fig:track_comp}
\end{figure}

\begin{figure}[t]
\centering
\includegraphics[width=3in,height=1.7in]{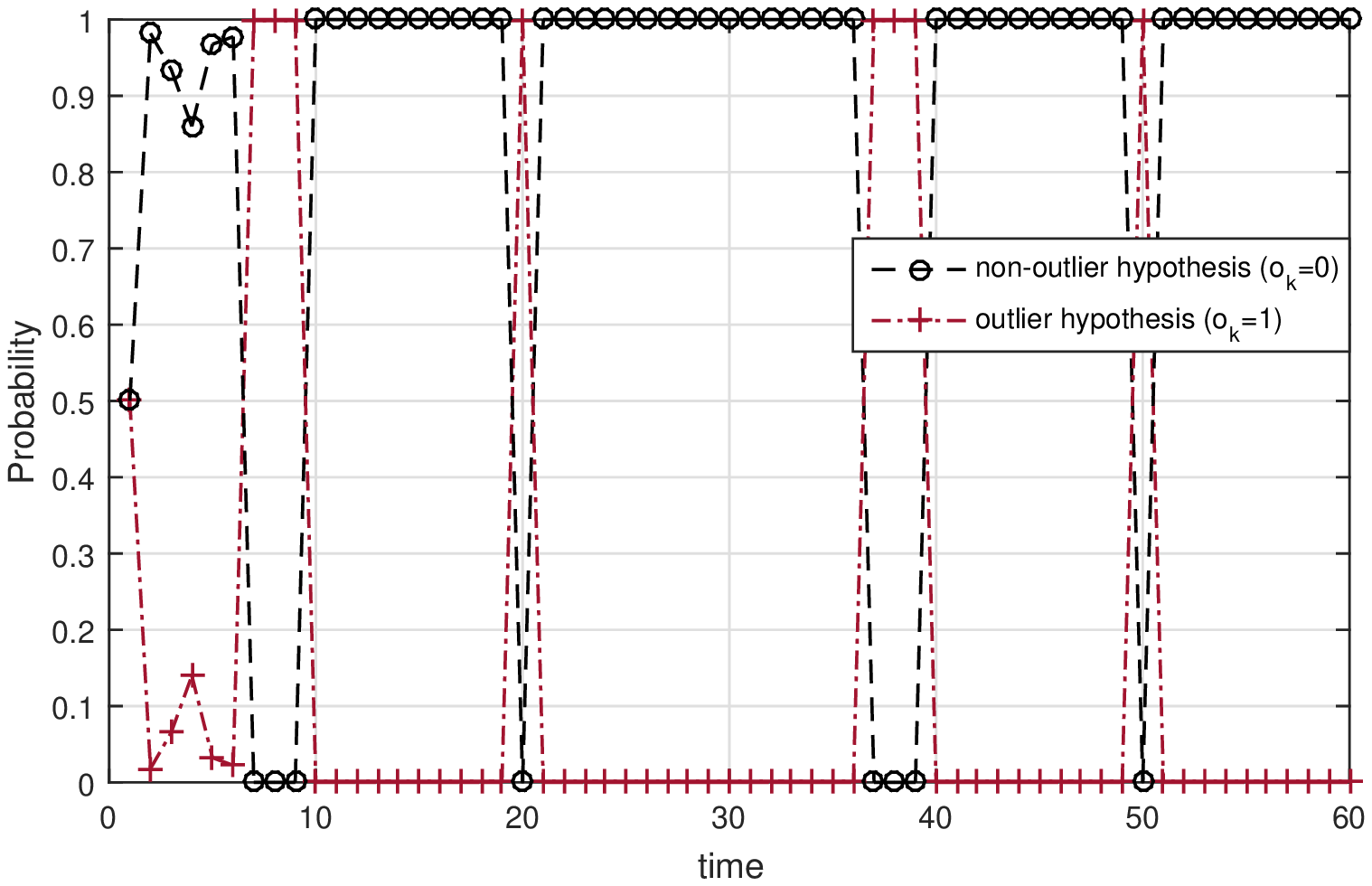}\\
\includegraphics[width=3in,height=1.7in]{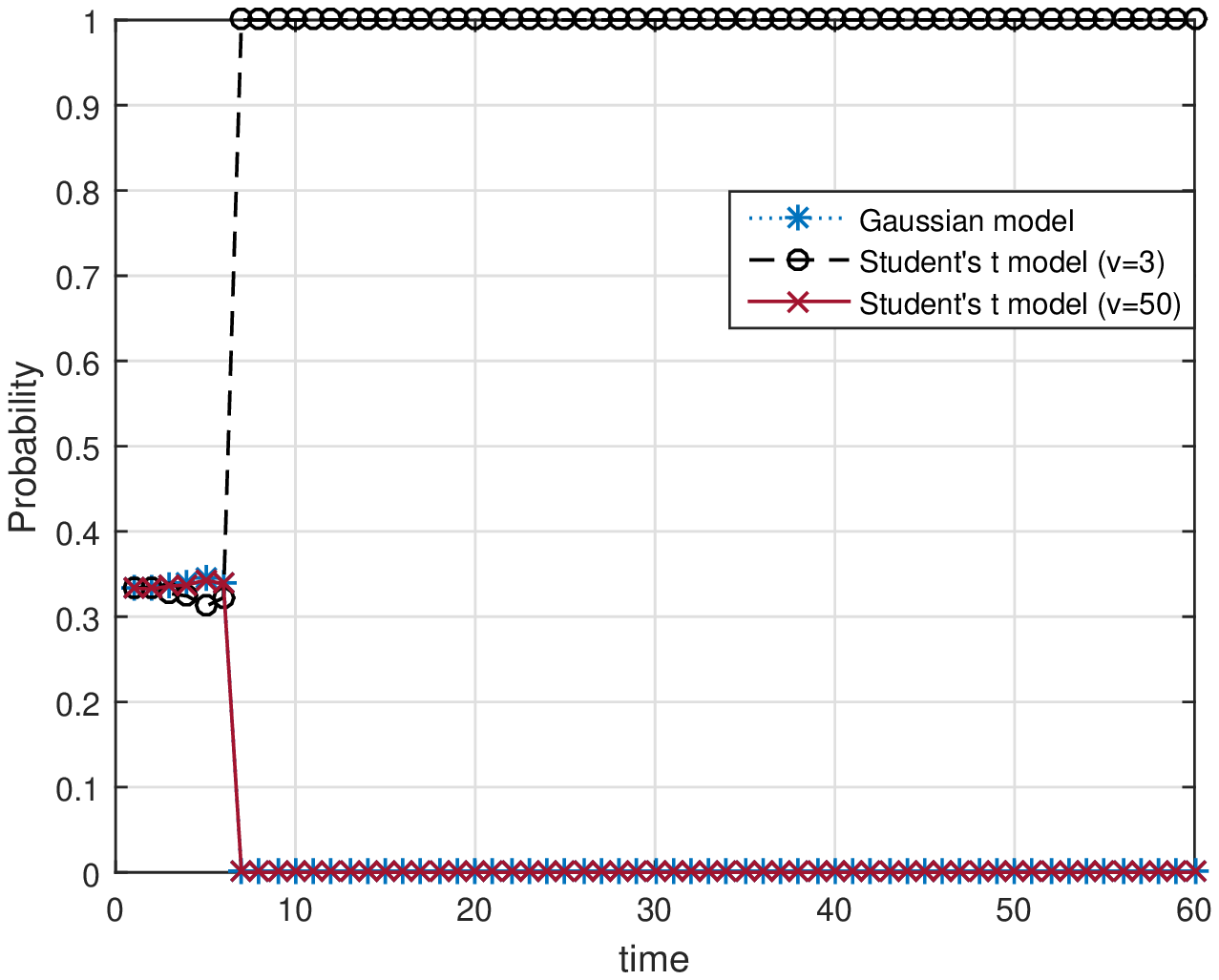}
\caption{The top panel shows the ILAPF yielded posterior probabilities of the outlier hypothesis (corresponding to $o_k=1$)
and the non-outlier hypothesis (corresponding to $o_k=0$).
The bottom panel shows the HMM-RPF \cite{liu2017robust} yielded posterior probabilities of the candidate models it employs and v
denotes the degree of freedom of a Student's t model. The real arrival time steps of the outliers are $7, 8, 9, 20, 37, 38, 39, 50$.}\label{fig:prob_o}
\end{figure}

Based on the above simulation setting, we first simulated a time-series
and then ran the ILAPF and HMM-RPF \cite{liu2017robust} to process the data, respectively.
The ILAPF is initialized with $\hat{lb}=0$, $\hat{ub}=70$ and $I=20$. The value range [0, 70] represents
a vague initial guess for the real value range [20, 30].
An $I$ value 20 is selected based on the simulation results as shown in Fig. \ref{fig:ore}.
The HMM-RPF algorithm in use is the same as that presented in \cite{liu2017robust}
with the forgetting factor $\alpha$ set at 0.9.
For both ILAPF and HMM-RPF, a particle size $N=200$ is used.
The state filtering and the posterior model/hypothesis inference results are shown in
Figs.\ref{fig:track_comp} and \ref{fig:prob_o}, respectively.
\begin{table}\centering\small
\caption{Execution time (in seconds), Mean and variance of the MSE calculated over 30 independent runs of each algorithm.}
\begin{tabular}{c||c||c|c}
\hline %
Algorithm & Time & \multicolumn{2}{c}{MSE} \\
& &mean&var \\\hline
ILAPF & 3.998 &0.365&0.007 \\\hline
HMM-RPF \cite{liu2017robust} &5.509&0.582&0.109\\\hline
\end{tabular}
\label{Table:convergence values}
\end{table}
\begin{table}[!htb]\centering\small
\caption{Mean and variance of the MSE calculated over 30 independent runs of ILAPF for 4 consecutive tasks.}
\begin{tabular}{c||c|c|c|c}
\hline
& Task 1&Task 2&Task 3&Task 4\\\hline
Mean of MSE& 0.365 &0.360& 0.333 &0.272 \\\hline
Variance of MSE&0.007&0.005&0.004&0.003\\\hline
\end{tabular}
\label{Table:mse_4_tasks}
\end{table}
\begin{figure}[!htb]
\centering
\includegraphics[width=3.4in,height=2.2in]{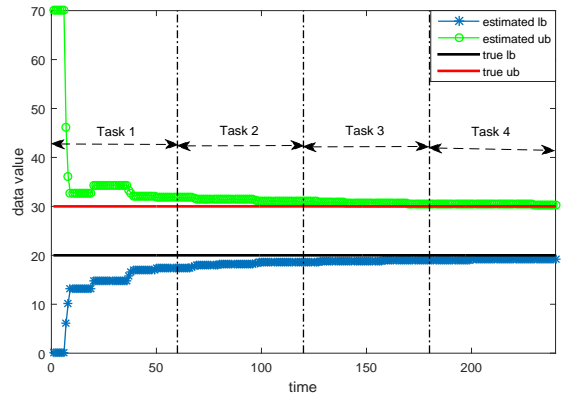}
\caption{The estimated lower and upper bounds of the outliers' value range over 4 consecutive state filtering tasks}\label{fig:bounds}
\end{figure}
We then conducted a Monte Carlo test for the involved algorithms.
We did the above experiment 30 times and then calculated the mean of execution time (in seconds),
the means and variances of the mean-square-error (MSE) of the state estimates over these experiments
for each algorithm. The result is summarized in Table 1.

Finally, we tested whether the incremental learning property of the ILAPF can provide benefits to implement a transfer learning
among related state filtering tasks. We simulated 4 consecutive tasks of state filtering in presence of outliers. For each task, the experimental setting is totally the same as presented in Subsection \ref{sec:setting}.
The ILAPF is run for each task one by one, and the learned value bounds of the outliers
from one task are used to initialize ILAPF for the subsequent task. The estimated bound over these 4 tasks
is shown in Fig.\ref{fig:bounds} and a quantitative MSE result
corresponding to 30 independent runs of the ILAPF for 4 consecutive tasks is presented in Table \ref{Table:mse_4_tasks}.
We see that, with aid of ILAPF, the learned information on outliers from one task can be transferred to
subsequent tasks, resulting in a continuous incremental performance gain, in terms of MSE, over tasks.
\section{Conclusions}
MMS is a powerful solution to address nonlinear state filtering problems in presence of model uncertainty.
The common practice to implement MMS is to specify a set of candidate models beforehand.
In this paper, we proposed a novel way to implement MMS in the context of nonlinear state filtering in presence of outliers.
Instead of specifying a set of candidate models beforehand, we select to learn a model to approximate the distribution of the outliers
in a sequential way. The resulting algorithm, ILAPF, is shown to be more accurate and faster than its competitor algorithm HMM-RPF \cite{liu2017robust}. Through simulations, we also show that the ILAPF algorithm makes transfer learning among related state filtering tasks possible.
\bibliographystyle{IEEEbib}
\bibliography{mybibfile}
\end{document}